\documentstyle[amssymb,12pt]{article}

\def\ker{{\rm ker}}

\def\otimesinf{\mathop{\otimes}}

\def\k{\ \Bbb K\ }
\def\Bbbone{\mbox{\rm 1\hspace {-.6em} l}}
\def\ea{{\'e}}
\def\eg{{\`e}}
\def\ec{{\^e}}
\def\ag{{\`a}}

\def\props{{propri\'et\'es}}
\def\hom{{\mbox{Hom}}}
\def\phom{{\mbox{\scriptsize Hom}}}
\def\der{{\mbox{\scriptsize Der}}}
\def\gder{{\mbox{Der}}}
\def\D{{\mbox{\scriptsize D}}}
\def\mer{{\mbox{\normalshape\scriptsize Der}}}

\newtheorem{proposition}{PROPOSITION}
\newtheorem{corollaire}{COROLLAIRE}

\begin{document}

\baselineskip=0.5cm
\begin{center}
{\large\bf D\ea rivations et calcul diff\ea rentiel non commutatif II}

\end{center}
\vspace{0.30cm}

\begin{center}
{Michel DUBOIS-VIOLETTE \&  Peter W. MICHOR}\\
\end{center}
\vspace{0.25cm}

\begin{small}
\noindent {\bf \it R\ea sum\'e} - Nous caract\ea risons la d\ea rivation
$d:A\rightarrow \Omega^1_{\der}(A)$ par une propri\'et\'e
universelle en introduisant une nouvelle classe de bimodules.
\end{small}

\vspace{0.30cm}
\begin{center}
{\bf D\ea rivations and noncommutative differential calculus II}\\
\end{center}

\begin{small}
\noindent {\bf \it Abstract} -{\it We characterize the derivation
$d:A\rightarrow \Omega^1_{\der}(A)$ by a universal
property introducing a new class of bimodules.}
\end{small}

\vspace{0.5cm}

{\it \bf Abridged English Version} - In this Note, $A$ denotes an
associative algebra over $\Bbb K=\Bbb R$ or $\Bbb C$ with a unit
$\Bbbone$. In the Note [5], a graded differential algebra
$\Omega_{\der}(A)$ with $\Omega^0_{\der}(A)=A$ was introduced with the notation
$\Omega_{\D}(A)$. It is one of the aims of this Note to show that, by
introducing a new category of bimodules, the derivation $d:A\rightarrow
\Omega^1_{\der}(A)$ is characterized by a universal property. Before
introducing this category of bimodules, we first consider a slightly
bigger category of bimodules : we call {\it
central bimodule} a bimodule such that the corresponding bimodule structure
over the center
$Z(A)$ of $A$ is induced by a structure of $Z(A)$-module. The category of
central bimodules is
a full subcategory of the category of all bimodules and we construct for
the corresponding embedding functor a left adjoint $M\mapsto M_Z$ and a
right adjoint $M\mapsto M^Z$. Consequently, {\it the category of central
bimodules is stable by taking subbimodules, quotient modules, arbitrary
projective limits and arbitrary inductive limits}. This category is
obviously stable by taking tensor products over $Z(A)$, and consequently
also over $A$. By applying the functor $M\mapsto M_Z$ to the universal
differential calculus on $d:A\rightarrow \Omega^1(A)$, we produce a
universal differential calculus for central bimodules, i.e. a derivation
$d_Z:A\rightarrow \Omega^1(A)_Z$ where $\Omega^1(A)_Z$ is central such
that any derivation of $A$ in a central bimodule factorizes through
$d_Z$ and a unique homomorphism from $\Omega^1(A)_Z$ in the bimodule. In
the case where $A$ is commutative, a central bimodule is simply a module and
 $\Omega^1(A)_Z$ reduces to the module
of K\"ahler $\Bbb K\,$-differentials $\Omega^1_{\Bbb K}(A)$.
We now introduce the appropriate category of bimodules to deal with
$d:A\rightarrow \Omega^1_{\der}(A)$. We call {\it diagonal bimodule} a
bimodule which is isomorphic to a subbimodule of an arbitrary product of $A$.
The category of diagonal bimodules is a full subcategory of
the category of bimodules and we produce a left adjoint $M\mapsto Diag\ (M)$
for the corresponding embedding functor. Consequently {\it the
category of Diagonal bimodules is stable by taking subbimodules and
arbitrary projective limits}. This category is also stable by taking
tensor products over $A$. By applying the functor $Diag$ to
$d:A\rightarrow \Omega^1(A)$, we obtain a universal differential
calculus $d:A\rightarrow Diag\ (\Omega^1(A))$ for the derivations of $A$ in
diagonal bimodules. Furthermore we show that $Diag\
(\Omega^1(A))=\Omega^1_{\der}(A)$ and that $d$ coincides with the
differential of $\Omega_{\der}(A)$. Finally, it is worth noticing that
when $A$ is commutative, a diagonal bimodule is simply a module such
that the canonical mapping in its bidual is injective. Thus the notions
of central bimodule and of diagonal bimodule are noncommutative
generalisations of the notion of module over a commutative algebra which
do not coincide with the notion of right (or left) module.

\vspace {0.5cm}
\noindent 1. INTRODUCTION ET PR\'ELIMINAIRES - Dans la Note [5], une
alg\eg bre diff\ea rentielle gradu\ea e $\Omega_{\der}(A)$ avec
$\Omega^0_{\der}(A)=A$ a \ea t\'e introduite (avec la notation
$\Omega_{\D}(A)$), $A$ \ea tant  une alg\eg bre unif\eg re. Le calcul diff\ea
rentiel $\Omega_{\der}(A)$, bas\'e sur
les d\ea rivations de $A$, a \'et\'e construit \`a partir du calcul
diff\ea rentiel universel $\Omega(A)$ [3], [7]. Du degr\'e z\'ero au degr\'e
un,
le calcul diff\'erentiel universel est caract\'eris\'e par le fait que
la d\'erivation $d:A\rightarrow \Omega^1(A)$ est universelle pour les
d\'erivations de $A$ \`a valeurs dans les bimodules. Dans cette Note,
nous montrons que la d\'erivation $d:A\rightarrow \Omega^1_{\der}(A)$
est universelle pour les d\'erivations de $A$ \`a valeurs dans une
cat\'egorie de bimodules que nous appelons bimodules diagonaux. Dans le
cas o\`u $A$ est commutative, un bimodule diagonal n'est autre que le
bimodule sous-jacent \`a un module tel que l'application canonique dans
son bidual est injective. Nous introduisons une autre cat\ea gorie de
bimodules, les bimodules centraux, qui est un peu plus grande que celle
des bimodules diagonaux et qui se r\ea duit  pour $A$ commutative \`a la
cat\ea gorie des modules. Cette cat\ea gorie poss\eg de aussi une d\ea
rivation universelle $d_Z:A\rightarrow \Omega^1(A)_Z$. Dans le cas o\`u
$A$ est commutative, $\Omega^1(A)_Z$ est le module des $\Bbb K$-diff\ea
rentielles de K\"ahler $\Omega^1_{\Bbb K} (A)$ et $d_Z$ est la $\Bbb
K$-diff\ea rentielle $d_{A/\Bbb K}$. Dans toute cette
Note $A$ d\ea signe une $\Bbb K\,$-alg\eg bre associative poss\ea dant une
unit\'e not\ea e $\Bbbone$ avec $\Bbb K = \Bbb R\ $ ou $\Bbb C\ $. Par un
bimodule sur $A$ ou plus simplement un bimodule, nous d\ea signerons
toujours un $(A,A)$-bimodule i.e. un $A\otimes A^{op}$-module. L'alg\eg
bre $A$ elle-m\ec me est canoniquement un bimodule. Si $M$ et $N$ sont
deux bimodules $\hom^A_A(M,N)$ d\ea signera l'espace des homomorphismes
de bimodules de $M$ dans $N$ ; on consid\eg rera $M\otimes N$ comme un
bimodule pour la structure d\ea finie par $a(m\otimes n) b=(am)\otimes
(nb)$, $a,b \in A$, $m\in M$ et $n\in N$.
$\displaystyle{M\otimesinf_A}N$ et $\displaystyle{M\otimesinf_{Z(A)} N}$
sont des bimodules quotients de $M\otimes N$. Soit $\Omega^1(A)$ le
noyau de la multiplication $m:A\otimes A\rightarrow A$, $m(a\otimes
b)=a\cdot b$, et soit $d:A\rightarrow \Omega^1(A)$ l'application lin\ea
aire d\ea finie par $da=\Bbbone \otimes a-a\otimes\Bbbone$~;
$\Omega^1(A)$ est un sous-bimodule de $A\otimes A$ et $d$ est une d\ea
rivation de $A$ \`a valeurs dans $\Omega^1(A)$. Le couple
$(\Omega^1(A),d)$ est caract\ea ris\'e, \`a un isomorphisme pr\`es, par
la propri\ea t\'e universelle suivante [1], [2] : {\it pour toute d\ea
rivation $\delta$ de $A$ \`a valeurs dans un bimodule $M$ il existe un
unique homomorphisme de bimodules} $i_\delta:\Omega^1(A)\rightarrow M$
{\it tel que} $\delta=i_\delta \circ d$, i.e. on a $\gder(A,M)\simeq
\hom^A_A(\Omega^1(A),M)$. Soit $\gder(A)(=\gder(A,A))$ l'alg\eg bre de Lie
des d\ea rivations de $A$ dans $A$ et soit $C^1(\gder(A),A)$ le bimodule
des applications lin\ea aires de $\gder(A)$ dans $A$. On d\ea signera
encore par $d$ la d\ea rivation de $A$ \`a valeurs dans $C^1(\gder(A),A)$
d\ea finie par $(da)(\delta)=\delta a$, $\forall \delta\in \gder(A)$ ; le
sous-bimodule de $C^1(\gder(A),A)$ engendr\'e par $dA$ sera not\'e
$\Omega^1_{\der} (A)$. L'homomorphisme canonique $i_d$ de $\Omega^1(A)$
dans $\Omega^1_{\der}(A)$ est surjectif, et son noyau
$F^1\Omega^1(A)$ est donn\'e par $F^1\Omega(A)=\cap\{\ker\ (i_\delta) \vert
\delta \in \gder (A)\}$, [5]. Il r\ea sulte de ce qui pr\ea c\eg de que
l'on a $\gder(A)\simeq
\hom^A_A(\Omega^1(A),A)\simeq\hom^A_A(\Omega^1_{\der}
(A),A)$. Dans la
r\ea f\ea rence [5], $\Omega^1_{\der}(A)$ a \ea t\'e introduit et \ea
tudi\'e avec la notation $\Omega^1_{\mbox{\D}}(A)$ ; nous avons remplac\'e le
``D" de [5] par ``$\gder$" afin d'\'eviter toute confusion avec [4]
o\`u la m\^eme notation $\Omega^1_{\mbox{\D}}(A)$ d\ea signe un autre
objet, (voir aussi [6] pour plus de d\ea tails sur $\Omega_{\der}(A)$). Dans
le cas o\`u $A$ est une alg\`ebre commutative, $\Omega^1(A)$ est un
id\ea al de $A\otimes A$ et le quotient $\Omega^1(A)/(\Omega^1(A))^2$
est canoniquement un $A$-module not\'e $\Omega^1_{\k}(A)$ et appel\'e {\it
module des} $\k$-{\it diff\ea rentielles de} $A$ ; l'image de
$d:A\rightarrow \Omega^1(A)$ par la projection canonique de
$\Omega^1(A)$ sur $\Omega^1_{\k}(A)$ est not\'ee $d_{A/\k}$, [1]. Le couple
$(\Omega^1_{\k}(A),d_{A/\k})$ est alors caract\ea ris\'e par la propri\ea
t\'e universelle suivante, [1] : {\it pour toute d\ea rivation $\delta$
de $A$ \`a valeurs dans un $A$-module $M$, il existe un unique
homomorphisme de $A$-module $i_\delta:\Omega^1_{\k}(A)\rightarrow  M$ tel
que $\delta=i_\delta \circ d_{A/\k}$}.\\

\noindent 2. BIMODULES CENTRAUX - Soit $M$ un bimodule sur $A$. Nous dirons
que $M$ est un {\it bimodule central} si pour tout \'el\'ement $z$
du centre $Z(A)$ de $A$ et pour tout $m\in M$ on a $zm=mz$. Il revient
au m\^eme de dire que la structure de bimodule de $M$ sur $Z(A)$ est
induite par une structure de $Z(A)$-module ou encore que $M$ est un
$A\displaystyle{\otimesinf_{Z(A)}}A^{op}$-module.

\noindent {\it Exemples.} -  1. Si $A$ est une alg\eg bre commutative,
i.e. $Z(A)=A$, un bimo-\- dule central sur $A$ n'est autre qu'un $A$-module
(pour la structure de
bimodule induite). - 2. Si le centre de $A$ est trivial, i.e. $Z(A)=\Bbb K
\Bbbone$, tout
bimodule est central.\\

\noindent On a les propri\ea t\ea s de stabilit{\ea}  suivantes:
{\it $($i$)$ tout sous-bimodule d'un bimodule central est central,
$($ii$)$ tout quotient d'un bimodule central est central,
$($iii$)$ tout produit de bimodules centraux est central,
$($iiii$)$ si $M$ et $N$ sont centraux alors
$\displaystyle{M\otimesinf_{Z(A)}}N$ est
central.}

\noindent Les propri\ea t\ea s de stabilit\'e ($i$), ($ii$) et ($iii$)
impliquent que {\it toute limite projective de bimodules centraux est
un bimodule central} et que {\it toute limite inductive de bimodules
centraux est un bimodule central}. Ces propri\ea t\ea s de stabilit\'e
sont reli\'ees \`a l'existence d'un foncteur covariant $M\mapsto M_Z$
adjoint \`a gauche et d'un foncteur covariant $M\mapsto M^Z$ adjoint \`a
droite du foncteur canonique $I_Z$ de la cat\ea gorie des bimodules
centraux dans celle des bimodules identifiant la cat\ea gorie des
bimodules centraux comme sous-cat\ea gorie pleine de la cat\ea gorie des
bimodules. Soit $M$ un bimodule quelconque (sur $A$) et soit $[Z(A),M]$ le
sous-bimodule de $M$ engendr\'e par les $zm-mz$ avec $z\in Z(A)$ et
$m\in M$. On d\ea signera par $M_Z$ le bimodule quotient $M/[Z(A),M]$ et
par $p_Z$ la projection canonique de $M$ sur $M_Z$. $M_Z$ est central et
tout homomorphisme de bimodules de $M$ dans un bimodule central est nul
sur $[Z(A),M]$. Soit d'autre part, $M^Z$ l'ensemble des \ea l\ea ments
$m\in M$ tels que $zm=mz$, $\forall z \in Z(A)$. $M^Z$ est un
sous-bimodule central de $M$, (le plus grand), on d\ea signera par $i^Z$
l'inclusion canonique de $M^Z$ dans $M$. Tout homomorphisme d'un
bimodule central dans $M$ a son image dans $M^Z$. Les couples
$(M_Z,p_Z)$ et $(M^Z,i^Z)$ sont donc caract\ea ris\'es \`a un
isomorphisme pr\`es par les propri\ea t\'es universelles suivantes.

\begin{proposition}.  $($i$)$ Pour tout homomorphisme de bimodules $\varphi :
M\rightarrow N$ de $M$ dans un bimodule central $N$, il existe un unique
homomorphisme de bimodules $\varphi_Z : M_Z \rightarrow N$ tel que
$\varphi = \varphi_Z \circ p_Z$. $($ii$)$ Pour tout homorphisme de
bimodules $\psi:N\rightarrow M$ d'un bimodule central $N$ dans $M$, il
existe un unique homomorphisme de bimodules $\psi^Z:N\rightarrow M^Z$
tel que $\psi = i^Z\circ \psi^Z$.
\end{proposition}
La partie ($i$) implique que le foncteur $M\mapsto M_Z$ de la cat\ea gorie
des bimodules dans la cat\ea gorie des bimodules centraux est adjoint
{\ag} gauche de $I_Z$, i.e. on a $\hom^A_A(M_Z,N)\simeq \hom^A_A(M,I_Z(N))$
pour $N$ central, et la partie ($ii$) implique que le foncteur $M\mapsto
M^Z$ est
adjoint \`a droite de $I_Z$, i.e. que l'on a $\hom^A_A(I_Z(N),M)\simeq
\hom^A_A(N,M^Z)$ pour $N$ central. Le foncteur
$M\mapsto M_Z$ est par cons\ea quent exact {\ag} droite et le foncteur
$M\mapsto
M^Z$ est exact \`a gauche.\\
On notera que $M$ est un bimodule central si et seulement si
$M=M_Z$ et que ceci est \ea quivalent \`a
$M=M^Z$. D'autre part, l'identit\'e $zm\otimes n
-m\otimes nz = mz\otimes n - m\otimes zn + (zm-mz)\otimes n+m\otimes (zn
-nz)$ implique que
si $M$ et $N$ sont des bimodules centraux alors $(M\otimes
N)_Z = \displaystyle{M \otimesinf_{Z(A)}}N$.

\noindent Consid\ea rons le bimodule $\Omega^1(A)$ et la (diff\ea rentielle)
d\ea
rivation universelle\linebreak[4] $d:A\rightarrow \Omega^1(A)$. Par composition
avec
la projection $p_Z:\Omega^1(A)\rightarrow \Omega^1(A)_Z$, on obtient une
d\ea rivation $d_Z:A\rightarrow \Omega^1(A)_Z$ ; $d_Z=p_Z\circ d$.

\begin{proposition}. Pour toute d\ea rivation $\delta:A\rightarrow M$ de
$A$ dans un bimodule central $M$, il existe un unique homomorphisme de
bimodules $i_\delta : \Omega^1(A)_Z\rightarrow M$ tel que l'on ait
$\delta = i_\delta \circ d_Z$.
\end{proposition}
Autrement dit, $d_Z:A\rightarrow \Omega^1(A)_Z$ est universelle pour les
d\ea rivations {\ag} valeurs dans les bimodules centraux. Cette
proposition est une cons\ea quence imm\ea diate de la proposition 2 et
de la propri\ea t{\ea} universelle de $d:A\rightarrow \Omega^1(A)$.

\begin{corollaire}. Si $A$ est une alg\eg bre commutative,
$\Omega^1(A)_Z$ s'identifie au $A$-module $\Omega^1_{\Bbb K}(A)$ des
$\Bbb K\,$-diff\ea rentielles (de K\"ahler) de $A$ et $d_Z$ est la\linebreak[4]
$\Bbb
K\,$-diff\ea rentielle $d_{A/\Bbb K}$.
\end{corollaire}

\noindent {\bf Remarque}. En utilisant l'exactitude {\ag} droite de
$M\rightarrow M_Z$, on d\ea duit de la suite exacte $0\rightarrow
\Omega^1(A)\buildrel j\over \longrightarrow A\otimes A \buildrel m\over
\longrightarrow A \rightarrow 0$ une suite exacte
$\Omega^1(A)_Z\buildrel {j^\sharp}\over \longrightarrow A
\displaystyle{\otimesinf_{Z(A)}} A\buildrel {m^\sharp}\over
\longrightarrow A \longrightarrow 0$. Dans le cas o\`u $A$ est
commutative, cette derni\eg re suite est une suite exacte de $A$-module
avec $\Omega^1(A)_Z=\Omega^1_{\Bbb K}(A)$ et
$A\displaystyle{\otimesinf_{Z(A)}} A=A\displaystyle{\otimesinf_A} A =
A$, mais $m^\sharp$ se r\ea duit alors {\ag} l'application identique de
$A$ dans $A$ et, par cons\ea quent, $\mbox{Im}(j^\sharp)=0$. Ceci montre
que $j^\sharp$ {\it n'est g\ea n\ea ralement pas injective}.\\

\noindent 3. BIMODULES DIAGONAUX - Soit $M$ un bimodule sur
$A$. Nous dirons que $M$ est un {\it  bimodule diagonal}
si $M$ est isomorphe {\ag} un
sous-bimodule de $A^I$ o\`u $I$ est un ensemble appropri{\ea} (d\ea
pendant de $M$). Tout bimodule diagonal est \ea videmment central.\\

\noindent {\it Exemples.} - 1. Si $A$ est une alg\eg bre commutative, un
bimodule diagonal n'est
autre qu'un A-module (pour la structure de bimodule induite) tel que
l'application canonique dans son bidual soit injective. - 2. Si $A$ est
l'alg\eg bre $M_n(\Bbb C\ )$ des matrices $n\times n$
complexes alors tout bimodule est diagonal.\\

\noindent On a les {\props} de stabilit{\ea} suivantes :
{\it $($i$)$ tout sous-bimodule d'un bimodule diagonal est diagonal, $($ii$)$
tout
produit de bimodules diagonaux est diagonal, $($iii$)$ si $M$ et $N$ sont
diagonaux alors
$M\displaystyle{\otimesinf_A} N$ est diagonal},
($A^I\displaystyle{\otimesinf_A} A^J$
est un sous-bimodule de $A^{I\times J}$).\\
\noindent Les propri\ea t\'es de stabilit\'e ($i$) et ($ii$)
impliquent que {\it toute limite projective de bimodules diagonaux est
un bimodule diagonal}. Ces propri\ea t\'es de stabilit\'e sont reli\'ees
\`a l'existence d'un foncteur {\it Diag} de la cat\ea gorie des
bimodules dans celle des bimodules diagonaux qui est adjoint \`a gauche
du foncteur canonique $I_{Diag}$ identifiant la cat\ea gorie des
bimodules diagonaux comme sous-cat\ea gorie pleine de la cat\ea gorie
des bimodules. Soit $M$ un bimodule quelconque et soit $c :M\rightarrow
A^{\phom^A_A(M,A)}$ l'homomorphisme canonique d\ea fini par
$c(m)=(\omega(m))_{\omega\in \phom^A_A(M,A)}$. Nous d\ea signerons par
$Diag(M)$ l'image de $c$. $Diag(M)$ est un bimodule diagonal.

\begin{proposition}. Pour tout homomorphisme de bimodules $\varphi :
M\rightarrow N$ de $M$ dans un bimodule diagonal $N$, il existe un
unique homomorphisme de bimodules
$\mbox{Diag}(\varphi):\mbox{Diag}(M)\rightarrow N$
tel que $\varphi=\mbox{Diag}(\varphi)\circ c$.
\end{proposition}
Puisque, par d\ea finition, un bimodule diagonal est un sous-bimodule de
$A^I$, il suffit de d\ea montrer le r\ea sultat pour $N=A^I$ ; cela
revient {\ag} le d\ea montrer pour $N=A$ ce qui est \ea vident d'apr\`es
les d\ea finitions. $\square$\\
La proposition 6 montre que le foncteur $I_{\rm Diag}$ admet le foncteur $Diag$
comme adjoint \`a
gauche, i.e. $\hom^A_A(Diag(M),N)\simeq \hom^A_A(M,I_
{Diag} (N))$. Le foncteur {\it Diag} est, par cons\ea quent, exact \`a
droite.\\
On notera que $M$ est un bimodule diagonal si et
seulement si $\hom^A_A(M,A)$ s\ea pare les points de $M$ et que ceci est
\ea quivalent \`a
$M\simeq Diag (M)$. D'autre part pour tout bimodule $M$, on a
$Diag (M)=Diag(M_Z)$, ce qui implique que si $M$ et $N$ sont des bimodules
centraux, (en particulier si ils sont diagonaux), on a
$Diag (M\otimes N) = Diag\displaystyle{(M\otimesinf_{Z(A)}N)}$.
\begin{proposition}. On a $Diag (\Omega^1(A))=\Omega^1_{\mer}(A)$.
\end{proposition}
$\Omega^1_{\der}(A)$ est, par construction, un sous-bimodule de
$A^{\der(A)}$ ; c'est donc un bimodule diagonal. D'autre part
$\hom^A_A(\Omega^1(A),A)\simeq \hom^A_A(\Omega^1_{\der}(A),A)$ implique
$\hom^A_A(\Omega^1(A),M)\simeq \hom^A_A(\Omega^1_{\der}(A),M)$
pour tout bimodule diagonal $M$, d'o\`u le r\ea sultat. $\square$
\begin{corollaire}
Pour toute d\ea rivation $\delta$ de $A$ \`a valeurs dans un bimodule
diagonal $M$, il existe un unique homomorphisme de bimodules $i_\delta
:\Omega^1_{\mer}(A)\rightarrow M$ tel que $\delta = i_\delta \circ d$.
\end{corollaire}
Autrement dit la d\ea rivation $d:A\rightarrow \Omega^1_{\der}(A)$ est
universelle pour les d\ea rivations \`a valeurs dans les bimodules
diagonaux. De m\^eme que $d:A\rightarrow\Omega^1(A)$ est le bloc de base
pour construire l'enveloppe diff\ea rentielle universelle $\Omega(A)$ de
$A$ [7], $d:A\rightarrow \Omega^1_{\der}(A)$ est le bloc de base pour
construire l'alg\eg bre diff\ea rentielle $\Omega_{\der}(A)$ introduite
dans [5] avec la notation $\Omega_{\D}(A)$.\\

\baselineskip=0.25cm
\begin{small}
\noindent R\'EF\'ERENCES BIBLIOGRAPHIQUES\\
\begin{description}
\item{[1]} N. BOURBAKI, {\sl Alg\eg bre} I, Chapitre III, Paris, Hermann 1970.
\item{[2]} H. CARTAN, S. EILENBERG, {\sl Homological algebra}, Princeton
University Press 1973.
\item{[3]} A. CONNES, Non-commutative differential geometry, {\sl Publi.
I.H.E.S.}, 62, 1986, p. 257.
\item{[4]} A. CONNES, {\sl Non commutative geometry}, Pr\ea pub.
I.H.E.S./m/93/54.
\item{[5]} M. DUBOIS-VIOLETTE, D\ea rivations et calcul diff\ea rentiel
non commutatif, {\sl C.R. Acad. Sci. Paris}, 307, S\ea rie I, 1988,
p.403-408.
\item{[6]} M. DUBOIS-VIOLETTE, P.W. MICHOR, The
Fr\"olicher-Nijenhuis bracket for derivation based non commutative
differential forms, to appear.
\item{[7]} M. KAROUBI, Homologie cyclique des groupes et des alg\eg
bres, {\sl C.R. Acad. Sci. Paris}, 297, S\ea rie I, 1983, p. 381-384.

\end{description}
\end{small}

\begin{flushright}
\makebox[3cm]{\hrulefill}

\parbox[]{8.5cm}{
\begin{description}
\item[]
{\scriptsize M.D-V. \it Laboratoire de Physique Th\ea orique et Hautes
Energies,\\
B\^at. 211, Universit\'e Paris XI, F-91405 Orsay Cedex}
\item[]
{\scriptsize P.W.M. \it  Erwin Schr\"odinger Institute of Mathematical
Physics,\\
Pasteurgasse 6/7, A-1090 Wien, Austria}
\end{description}}
\end{flushright}

\end{document}